# MHSP in reversed-biased operation mode for ion blocking in gas-avalanche multipliers

J.F.C.A. Veloso[a,b], F. Amaro[a,d], J.M. Maia[a,c,d], A.V. Lyashenko[d],
A Breskin[d], R. Chechik[d], JMF dos Santos[a], O. Bouianov[e], M. Bouianov[f]

[a]*Physics Dept., University of Coimbra, 3004-516 Coimbra, Portugal*
[b]*Physics Dept., University of Aveiro, 3810-193 Aveiro, Portugal*
[c]*Physics Dept., University of Beira Interior, 6200-001 Covilhã, Portugal*
[d]*Dept. of Particle Physics, The Weizmann Institute of Science, 76100 Rehovot, Israel*
[e]*Lab. for Nuclear Science, Massachusetts Institute of Technology, Cambridge, MA 02139, USA*
[f]*CSC – Scientific Computing Ltd, P.O. Box 405, FIN-02101 Espoo, Finland*

**Abstract**

We present recent results on the operation of gas-avalanche detectors comprising a cascade of gas electron multipliers (GEMs) and Micro-Hole and Strip Plates (MHSPs) multiplier operated in reversed-bias (R-MHSP) mode. The operation mechanism of the R-MHSP is explained and its potential contribution to ion-backflow (IBF) reduction is demonstrated. IBF values of $4\times10^{-3}$ were obtained in cascaded R-MHSP and GEM multipliers at gains of about $10^4$, though at the expense of reduced effective gain in the first R-MHSP multiplier in the cascade.

**Corresponding author**: J.M.F. dos Santos (jmf@gian.fis.uc.pt)

Tel: ++ 351 239 410667; Fax: ++ 351239 838850

# 1. Introduction

The present study investigates the possibility of reducing the yield of back-flowing avalanche-ions in gaseous detectors. The avalanche-induced ions are responsible for secondary effects, limiting the multiplier's gain and lifetime, and their suppression has been realised to be a key issue for the detector's performance.

In gaseous photomultipliers (GPMs) [1], ions flowing back and impinging on the photocathode (PC) induce its physical and chemical aging. The ions also induce secondary electron emission that results in excessive secondary avalanches known as ion-feedback; these cause gain limitations and localization deterioration. The problem is particularly acute with visible-sensitive GPMs due to the high secondary emission probability of visible sensitive PCs.

In time projection chambers (TPCs), ions flowing back from the multiplier into the conversion/drift region locally modify the electric field, resulting in dynamic track distortions [2]. This seriously affects the tracking properties of TPCs in high-multiplicity experiments, e.g. in relativistic heavy-ion physics applications. In both GPM and TPC cases, the ion back-flow should be suppressed to the sub per-mil scale or roughly $G^{-1}$, G being the multiplier's gain.

***The ion backflow*** (IBF) is defined as the fraction of total avalanche-generated ions reaching the PC in a GPM or the primary-ionization drift-volume in a tracking detector or in a TPC. The IBF depends on the multiplier's geometry, on the type of gas and pressure and on the electric fields.

In most commonly used gaseous detectors' configurations, e.g. Multi-wire proportional chambers (MWPCs), parallel-plate counters, resistive-plate chambers and others, almost all avalanche ions flow back to the cathode or to the collection region preceding the multiplier. Predictions were made to block ions in multi-grid avalanche detectors, by alternating high- and low-field regions [3]. Low IBF values of the order of $2\times10^{-3}$ were also predicted in Micromegas detectors optimized for TPC applications, due to the high fields ratio on both sides of the micromesh, resulting in its low ion-transparency [4]. The IBF can be reduced by many orders of magnitude by incorporating a pulsed ion-gate electrode that takes advantage of the natural delay in the ion arrival following an avalanche and blocks them, though at the expense of a considerable dead time; this common practice in TPCs [5] was successfully applied to GPMs incorporating UV [6] or visible-sensitive [7,8] PCs coupled to cascaded Gas Electron Multipliers (GEM [9]).

Detectors incorporating high-gain cascaded GEMs offer many attractive properties, e.g. complete screening of photon-mediated secondary processes and secondary effects; they became useful detection tools in various fields [10]. Intuitively such multi-element structures, with their alternating high/low electric fields in the holes and between the elements, were expected to efficiently block the ions. However, it was found [6,11, 12] that though the IBF is indeed a function of these fields ratio, a large fraction of the ions return to the cathode, following the same path as the electrons but in an opposite direction; this is due to the strong focussing of charges into the GEM holes, under the high-gain operation conditions of multi-GEM cascades. The IBF results obtained in numerous studies in cascaded multipliers were so far insufficient for the operation of TPCs at high rates and for GPMs with PCs sensitive in the visible spectral range. (A concise discussion on the subject is given elsewhere [7,13]). In cascaded GEMs, IBF values between a fraction of a % to a few % were reached, at best, depending primarily on the electric field above the first multiplying element ($E_{drift}$), on the total gain and on the hole-geometry of the GEMs in the cascade [12,14-16]. The above quoted values are for $E_{drift}$ values of the order of 0.1kV/cm, as in TPCs. In GPMs with a semitransparent PC, the field $E_{drift}$ at the PC surface must be higher, i.e. above 0.5 kV/cm, to ensure an efficient photoelectron extraction into the gas [1]; therefore, the IBF in multi-GEM GPMs could be reduced at best to levels of ~10%-20% at a gain of $10^5$ [6,13].

A significant step forward was the introduction of the Microhole & Strip Plate (MHSP [16]) electrode within the GEM cascade. This is a GEM-like electrode with extra anode strips patterned at its bottom. The MHSP is polarized such that a high field is established within the holes, as in GEM, and another strong filed is established at the anode strips. This results in two successive multiplication stages - in the holes and at the anode-strips. In the second multiplication step at the anode strips, the ions' and electron' paths split, and a significant part of the ions is collected on the neighbouring cathode strips and on the cathode plane placed below the MHSP. Thus, when the MHSP was used as a last element in the cascade, following 3 GEMs, a further reduction of the IBF to ~2-3%, at effective gains of $10^5$-$10^6$, was reported [17,18].

In an attempt to further reduce the IBF value, we recently implemented the reversed-bias MHSP (R-MHSP) as a first multiplying element in a GEM cascade. In this suggested mode [19], the extra patterned strips at the bottom of the electrode are biased as cathodes. Consequently, the avalanche occurs only within the GEM-like holes, while the extra cathode strips can attract a fraction of the up-flowing ions originating from avalanches in subsequent multiplying elements (Fig.1a). Typical avalanche-ion paths simulations in such a detector configuration, using MAXWELL[1] and GARFIELD [20,21] software packages are shown in fig.1b. The idea of splitting the ions and electrons path is maintained, as in MHSP. However, the extra cathode strips affect also the electrons paths, and the ion trapping occurs at the cost of a drop in the number of electrons transferred from the R-MHSP to the subsequent element. A careful optimization of the voltages applied to the different electrodes of the R-MHSP allows reaching an effective ion-backflow reduction, but a compromise regarding its effective gain and the resulting efficiency of electron transfer to subsequent multiplication elements must be taken into account.

## 2. Methodology and results

The measurements presented in this work were done in atmospheric pressure Ar/5%CH$_4$ mixture, in a gas-flow mode. GPMs with a semitransparent CsI PC coupled to a single- or to double- R-MHSPs followed by two GEMs were investigated. The ion-blocking capability of the R-MHSP itself was investigated using the set-up depicted in Fig.2. Photoelectrons emitted from the semi-transparent CsI PC irradiated with UV photons, are multiplied at the anode plane of the MWPC, positioned below the MHSP. This MWPC is used exclusively as an ion-source: while the electrons are collected at the anode wires, the ions flowing in the direction of $M_2$ are attracted by the field $E_{trans}$ towards the R-MHSP. We defined *the ion transparency* of the R-MHSP as the fraction of ions crossing the R-MHSP, namely the ratio between the current of ions exiting the holes, $I_{out}$, and the current of ions reaching the R-MHSP, $I_{in}$. $I_{in}$ was measured, for different $E_{trans}$, by interconnecting all the R-MHSP electrodes while maintaining the electric field $E_{trans}$ between $M_2$ and the R-MHSP bottom (Fig.2). $I_{out}$ was measured over a range of values of all potentials. Fig.3 depicts the ion transparency as a function of $V_{A-C}$, for different values of $V_{A-T}$ (Fig.3a) and different transfer fields, $E_{trans}$ (Fig.3b). For $V_{A-C}=0$, i.e. a R-MHSP operated in a GEM-mode, the ion transparency is around $0.2 - 0.3$; it decreases by more than factor 100 with increasing $V_{A-C}$, demonstrating the principle of ion blocking.

---

[1] Maxwell 3D Field Simulator, Ansoft Corporation.

However, the biasing of the cathode strips on the R-MHSP also affects the electron transmission and thus, the effective gain of the multiplier. The R-MHSP effective gain is defined as the ratio between the total electron charge transferred to the elements below the R-MHSP and the primary-electrons charge, originated in the drift region above it (see Fig1a). The electron charge transferred to the elements below the R-MHSP was measured by interconnecting all the electrodes of those elements and reading their current (see Fig.1a). The primary electrons charge was assessed by measuring the primary PC current, $I_{PC0}$, when only $E_{Drift}$ is established, and the three electrodes of the R-MHSP are inter-connected. Fig.4 depicts the R-MHSP effective gain as a function of $V_{A-C}$, for different values of $V_{A-T}$ (Fig.4a) and of the transfer field, $E_{trans}$ (Fig.4b). The drift field was set in all measurements to 0.5 kV/cm. As shown in Fig. 4, the effective gain decreases with increasing $V_{A-C}$ due to a considerable trapping of the avalanche electrons by the anode strips; it therefore limits the applicability of the reversed biasing. For example, for a $V_{A-C}$ value of 140V, needed for reducing the ion transparency to values below 1% (Fig. 3a), an effective gain of ~2 at best was reached in our present conditions (Fig.4a), which is of a significant drawback.

The effective gain of the first element in a cascade is a very important parameter. Regardless of the total gain of the cascaded multiplier, the gain in the first element defines the detection efficiency of the whole detector to single electrons in a GPM and the energy resolution (electron statistics) in the case of ionization measurements in a TPC. Effective gains >10 and total gains > $10^4$ are requested to assure full single-electron detection efficiency.

The effect of the IBF reduction with a R-MHSP incorporated in a cascaded multiplier was studied in a GPM comprising a semitransparent CsI PC coupled to a R-MHSP followed by two GEMs (Fig.5). Currents measured on different electrodes provided the ion-flow yields, normalized to the avalanche charge. The transfer fields between the R-MHSP and the first GEM and between the GEMs were fixed at 2 kV/cm; equal voltage differences, $V_{GEM}$, were applied across both GEMs. The drift field above the R-MHSP was set to 0.5 kV/cm.

The fraction of ion back-flow to the drift region, $IBF_{Drift}$, relevant to TPCs and to GPMs with semitransparent PCs, is derived from the avalanche-induced currents measured on the various electrodes:

$$IBF_{Drift} = \frac{(I_{PC} - I_{PC_0})}{(I_{BOT} + I_M)} \qquad (1)$$

where, $I_{PC}$ is the ion current induced on the PC, $I_{PC0}$ is the primary photocurrent, $I_{BOT}$ is the electron charge collected at the bottom electrode of the last GEM and $I_M$ is the electron current collected at the anode mesh placed below the last GEM.

The values for $IBF_{Drift}$ are presented in Fig.6 as a function of $V_{A-C}$ (Fig. 6a) and as a function of the total effective gain of the cascaded multiplier (Fig. 6b), for different $V_{GEM}$ and $V_{A-T}$ values. The effective gain is derived from the ratio between the current on the electrodes below the last GEM, $I_{Bot}+I_M$, and the primary photoelectron current, $I_{PC0}$. For $V_{A-T}$ voltages about 300 V, the best $IBF_{Drift}$ value of ~0.008, was obtained for a total effective gain of about $5\times10^3$ and for a reversed bias around 150 V. This represents a reduction by a factor of ~5 in the $IBF_{Drift}$ as compared to the R-MHSP operated in a GEM-mode (i.e. with $V_{A-C} = 0$), as sown in Fig.6a. Our results are in agreement with those obtained by Roth [19]. The minimum in the $IBF_{Drift}$ graphs is a result of competing effects of the various fields. It is obvious that ion deviation and trapping is first improving with increasing ratio $V_{A-C}/V_{A-T}$, but with this ratio being too high there are no more electrons transferred to the next elements and IBF starts increasing. The trend is that higher $V_{GEM}$ (i.e. higher total gain) and lower $V_{A-T}$ (i.e. more effective ion deviation) are pushing the IBF minimum to lower values. Unfortunately, we found that for the above conditions ($V_{A-T} = 300$ V; $V_{A-C} = 150$V) of minimal $IBF_{Drift}$, the gain of the R-MHSP is less than 1 (Fig.4a); this configuration with the presently applied potentials is therefore not applicable, as discussed above.

The above result indicates that we cannot afford to trap all ions on the first element, because at the same time we "kill" all the electrons. Better results were obtained in a four-element cascade of a double R-MHSP and a double GEM, shown in Fig.7. This arrangement allows us to maintain sufficient gain in the first R-MHSP and improve ion trapping not by pushing $V_{A-C}$ too much but rather by doing it in two steps. Both GEMs were polarized with a resistive network, maintaining proportionality between $V_{GEM}$ and the transfer voltage applied between both GEMs; therefore, the transfer field between the two GEMs was not constant, e.g. being 2 kV/cm for a $V_{GEM}$ voltage of 400 V. The induction field between the mesh and the last GEM had always the same value as the transfer field between the two GEMs. The transfer fields between both R-MHSPs and between the second R-MHSP and the first GEM were set to 2 kV/cm; the drift field was set to 0.5 kV/cm; equal $V_{A-C}$ and $V_{A-T}$ values were set on both R-MHSPs.

Similar current measurements and IBF calculations, as described above, were done. The results obtained for the $IBF_{Drift}$ are presented in Fig. 8 as a function of $V_{A-C}$ (Fig.8a) and of the total effective gain (Fig.8b), for different $V_{GEM}$ and $V_{A-T}$ values. They show that the additional R-MHSP further reduces the IBF value, at higher gains of the first R-MHSP. As expected, the minimum value for IBF is reached at lower $V_{A-C}$ values, around 50-60 V. This is of an advantage, because it results in a higher effective gain of the R-MHSP as compared to the effective gain at $V_{A-C} \sim 150$ V (see Fig. 4). As shown in Fig.4a, the R-MHSP gain is only reduced by a factor of ~2.5 when $V_{A-C}$ increases from 0 to 60 V, compared to the ~40-fold loss observed when $V_{A-C}$ increases from 0 to 150 V. For the above conditions, the R-MHSP effective gain is ~6 for the applied $V_{A-T}=300V$; as reflected from fig. 4a, gains >10 can be reached already for $V_{A-T}$ values of ~320V, which in principle should be possible but could not be set in this work due to defects in the electrode and consequent electrical instabilities. The double R-MHSP & double GEM cascade operated with $V_{A-T}$ of 300V and $V_{A-C}$ of 60 V yielded, at best, $IBF_{Drift}$ values of about 0.004 and 0.01 for gains around $10^4$ and $10^5$, respectively.

**3. Summary and conclusions**

The capability of the MHSP operated in reversed bias mode (R-MHSP) to reduce the ion backflow (IBF) in avalanche detectors was demonstrated. $IBF_{Drift}$ values of about 0.008 were reached at total gains of about 5 x $10^3$ in a gaseous photomultiplier (GPM) with a semitransparent CsI photocathode (PC) coupled to a cascaded R-MHSP plus double-GEM multiplier; the potentials between the R-MHSP strips and across the hole were $V_{A-C}$=150V and $V_{A-T}$=300V, respectively. However, we have shown that in these conditions, the gain of the R-MHSP is below 1. This low gain, not discussed in Ref. [19], is unusable; effective gains of at least 10 are needed for the first element in the cascade in order not to loose single-photon events in a GPM and not to affect the primary-electron statistics in a TPC.

Better results were demonstrated in a detector configuration of a semitransparent PC followed by two R-MHSPs and a double-GEM. An $IBF_{Drift}$ value of 0.004 was reached at total a gain of $10^4$. The applied potentials were: $V_{A-C}$=60V between the R-MHSP strips, $V_{A-T}$=300V across the hole, resulting in a R-MHSP effective gain of ~6.

This $IBF_{Drift}$ value is about 5 times better compared to that of a triple-GEM&MHSP cascade and about 50 times better than that of a quadruple-GEM.

However, it is still more than one order of magnitude above the desired value of $G^{-1}$. Moreover, the effective gain of the first element is still rather low (~6-10), and so is the total gain of the cascade (~$10^4$), which may be insufficient for the efficient detection of single photoelectrons.

Our study reveals the potential of this approach and we are confident that by further increasing the multiplication in the elements below the double R-MHSP, e.g. by increasing $V_{GEM}$, adding another GEM, or using a THGEM [22] with its 10 time higher gain, better results may be demonstrated. It should be noted that the use of lower values of $E_{Drift}$, e.g. 0.1 kV/cm, would further reduce $IBF_{Drift}$ since more ions will be collected at the R-MHSP top electrode [14].

Further studies are in course, with other multiplier configurations. These include the use of additional GEM and MHSP elements, better quality MHSP electrodes, optimization of the various electric fields etc. Preliminary results are summarized in [7]. These investigations are expected to further reduce the IBF values in gas-avalanche multipliers.

**Acknowledgements**

This work was supported in part by Project POCTI/FP/FNU/50212/03 through FEDER and FCT (Lisbon) programs and by the Israel Science Foundation project 151/01. A. Breskin is the W.P. Reuther Professor of Research in peaceful use of atomic energy.

**Figure Captions**

Fig.1 – Schematic of a R-MHSP operation principle (*a*) and of the simulated ion-drift paths in a R-MHSP (*b*).

Fig.2 – Schematic of a photon detector with a reflective CsI photocathode and multiwire chamber, used as an ion source for the study of the ion transparency in a R-MHSP.

Fig.3 – R-MHSP ion transparency as a function of $V_{A-C}$ measured in the detector shown in Fig.2, at atmospheric pressure of Ar/5%CH$_4$: (*a*) for different values of $V_{A-T}$ and $E_{trans}$= 2.0 kV/cm; (*b*) for different values of the $E_{trans}$, and $V_{A-T}$ = 300 V. $E_{drift}$=0.5kV/cm in all measurements.

Fig.4 – R-MHSP effective gain as a function of $V_{A-C}$ measured in the detector shown in Fig.1*a*, at atmospheric pressure of Ar/5%CH$_4$: (*a*) for different values of $V_{C-T}$ and $E_{trans}$= 2.0 kV/cm; (*b*) for different values of $E_{trans}$ and $V_{C-T}$ = 300 V. $E_{drift}$=0.5 kV/cm in all measurements.

Fig.5 – Experimental setup for the study of the ion backflow in a gaseous photomultiplier combining a semitransparent photocathode and a R-MHSP cascaded with a double-GEM.

Fig.6 – Ion backflow to the drift region of the detector shown in fig. 5 as a function of $V_{A-C}$ (*a*) and of the total effective gain (*b*), for different $V_{A-T}$ and $V_{GEM}$ values; $E_{drift}$=0.5kV/cm; Ar/5%CH$_4$ at atmospheric pressure.

Fig.7 - Experimental setup for the study of ion backflow in a gaseous photomultiplier combining a semitransparent photocathode and a double-R-MHSP cascaded with a double-GEM.

Fig.8 - Ion backflow to the drift region of the detector shown in fig. 7 as a function of $V_{A-C}$ (*a*), and of the total effective gain (*b*), for different $V_{A-T}$ and $V_{GEM}$ values; $E_{drift}$=0.5kV/cm; Ar/5%CH$_4$ at atmospheric pressure.

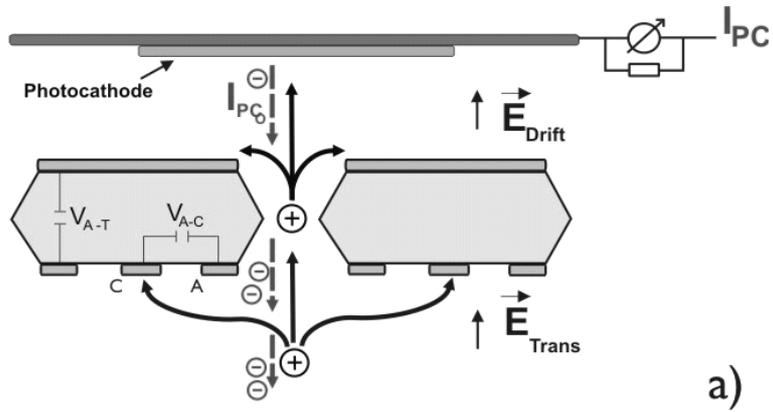

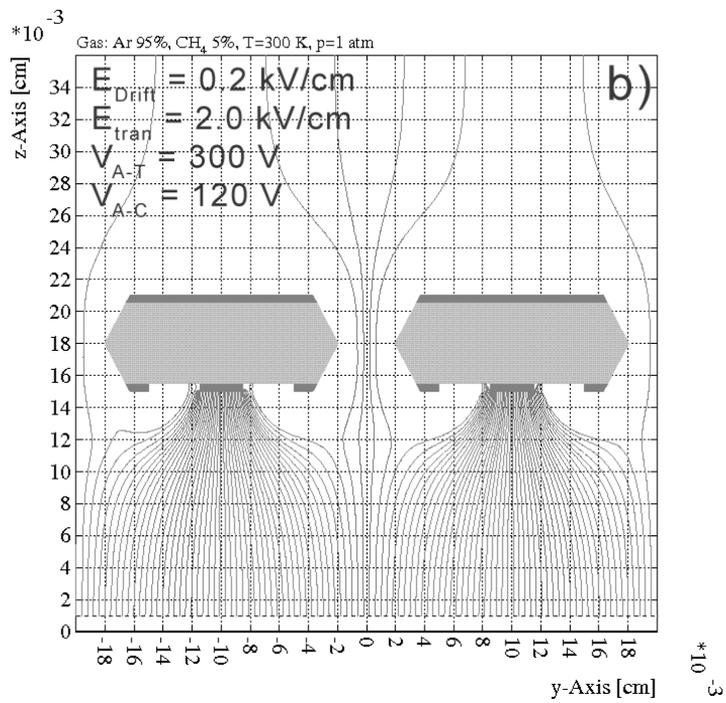

Fig.1 – Schematic of a R-MHSP operation principle (*a*) and of the simulated ion-drift paths in a R-MHSP (*b*).

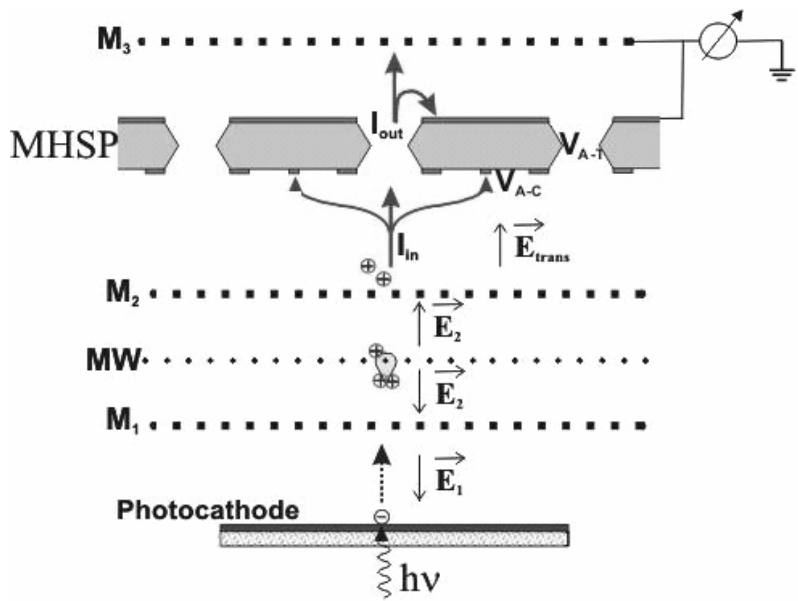

Fig.2 – Schematic of a photon detector with a reflective CsI photocathode and multiwire chamber, used as an ion source for the study of the ion transparency in a R-MHSP.

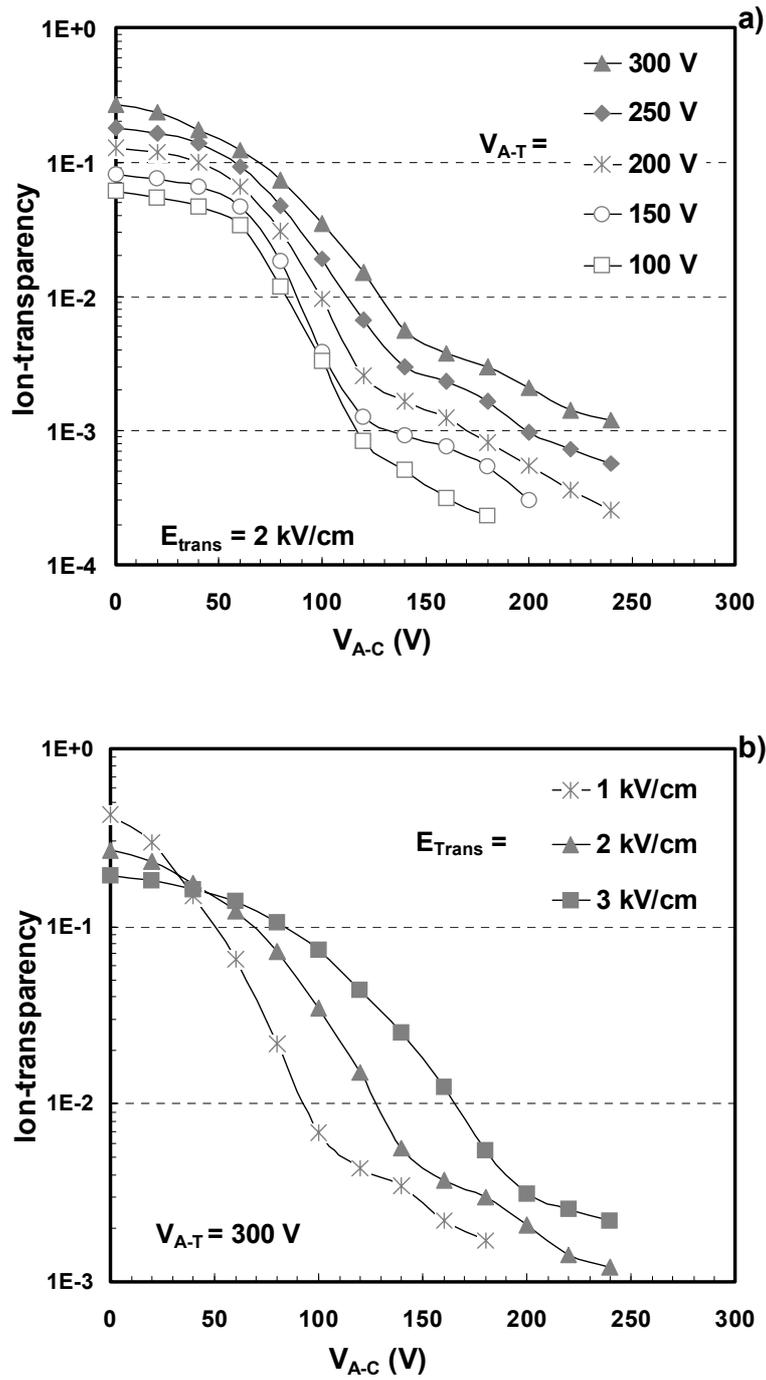

Fig.3 – the R-MHSP ion transparency as a function of $V_{A-C}$ measured in the detector shown in Fig.2, at atmospheric pressure of Ar/5%CH$_4$: (*a*) for different values of $V_{A-T}$ and $E_{trans}$= 2.0 kV/cm; (*b*) for different values of the $E_{trans}$, and $V_{A-T}$ = 300 V. $E_{drift}$=0.5kV/cm in all measurements.

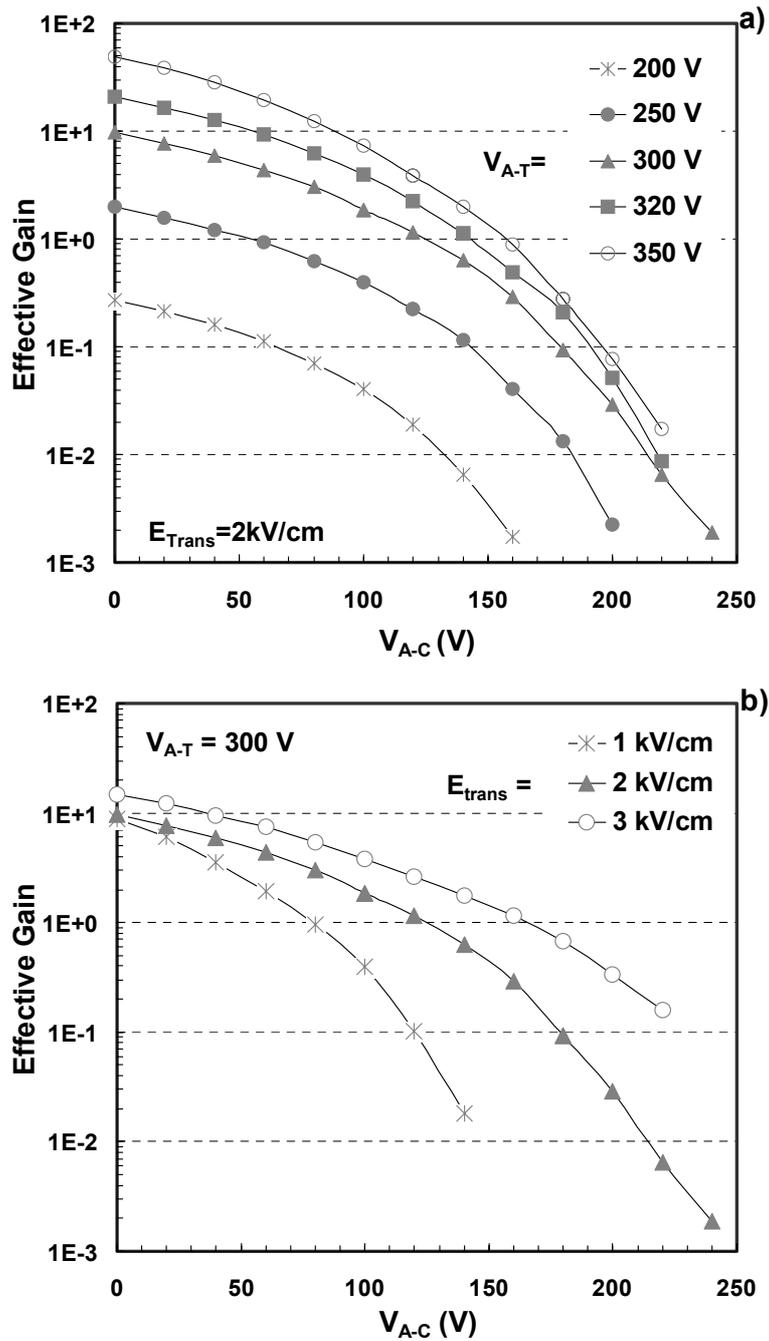

Fig.4 – R-MHSP effective gain as a function of $V_{A-C}$ measured in the detector shown in Fig.1*a*, at atmospheric pressure of Ar/5%CH$_4$: (*a*) for different values of $V_{C-T}$ and $E_{trans}$= 2.0 kV/cm; (*b*) for different values of $E_{trans}$ and $V_{C-T}$ = 300 V. $E_{drift}$=0.5 kV/cm in all measurements.

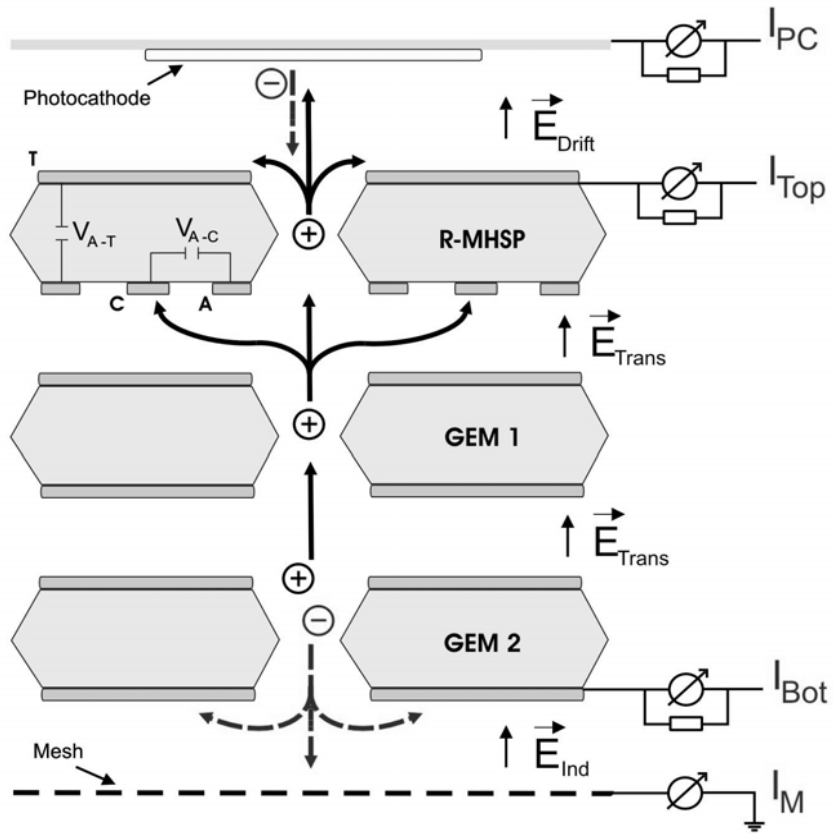

Fig.5 – Experimental setup for the study of the ion backflow in a gaseous photomultiplier combining a semitransparent photocathode and a R-MHSP cascaded with a double-GEM.

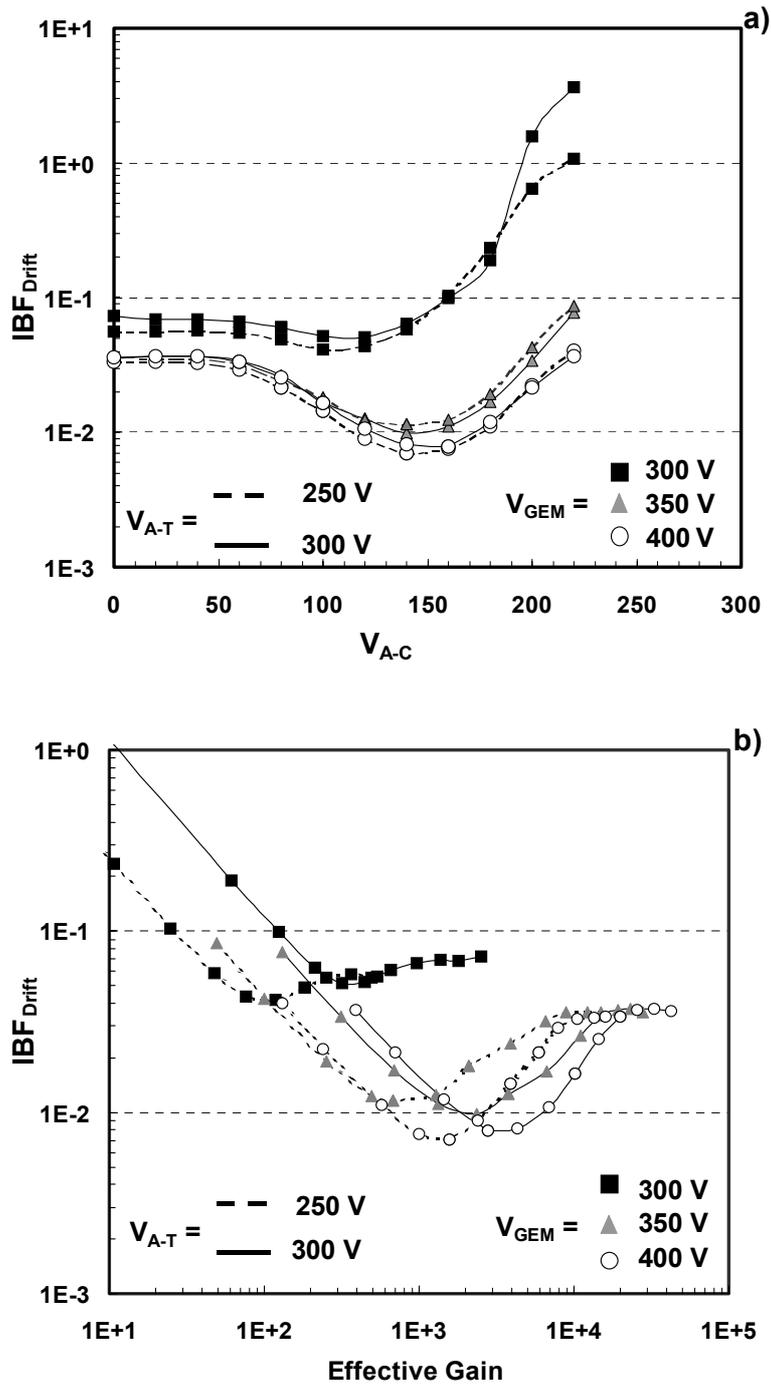

Fig.6 – Ion backflow to the drift region of the detector shown in fig. 5 as a function of $V_{A-C}$ (*a*) and of the total effective gain (*b*), for different $V_{A-T}$ and $V_{GEM}$ values; $E_{drift}$=0.5kV/cm; Ar/5%CH$_4$ at atmospheric pressure.

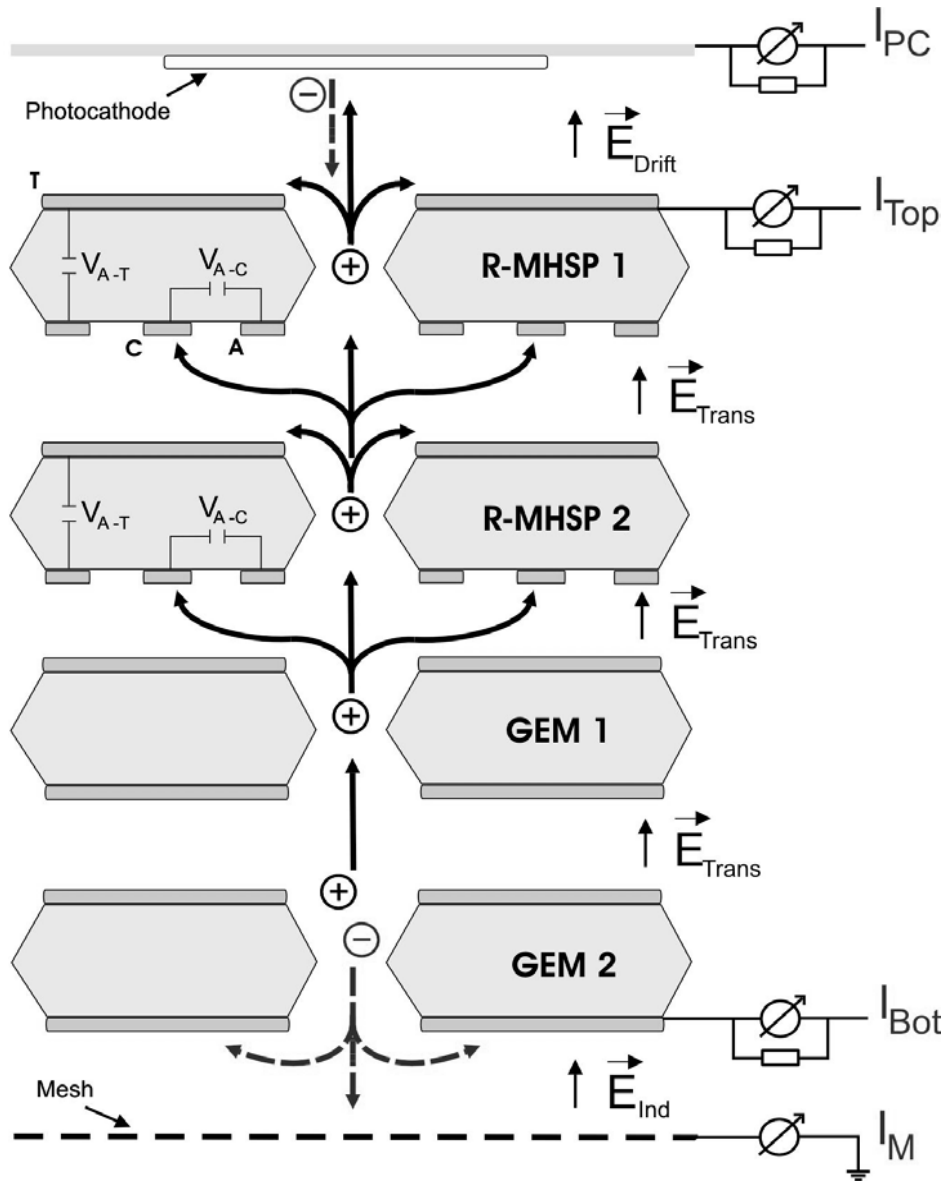

Fig.7 - Experimental setup for the study of ion backflow in a gaseous photomultiplier combining a semitransparent photocathode and a double-R-MHSP cascaded with a double-GEM.

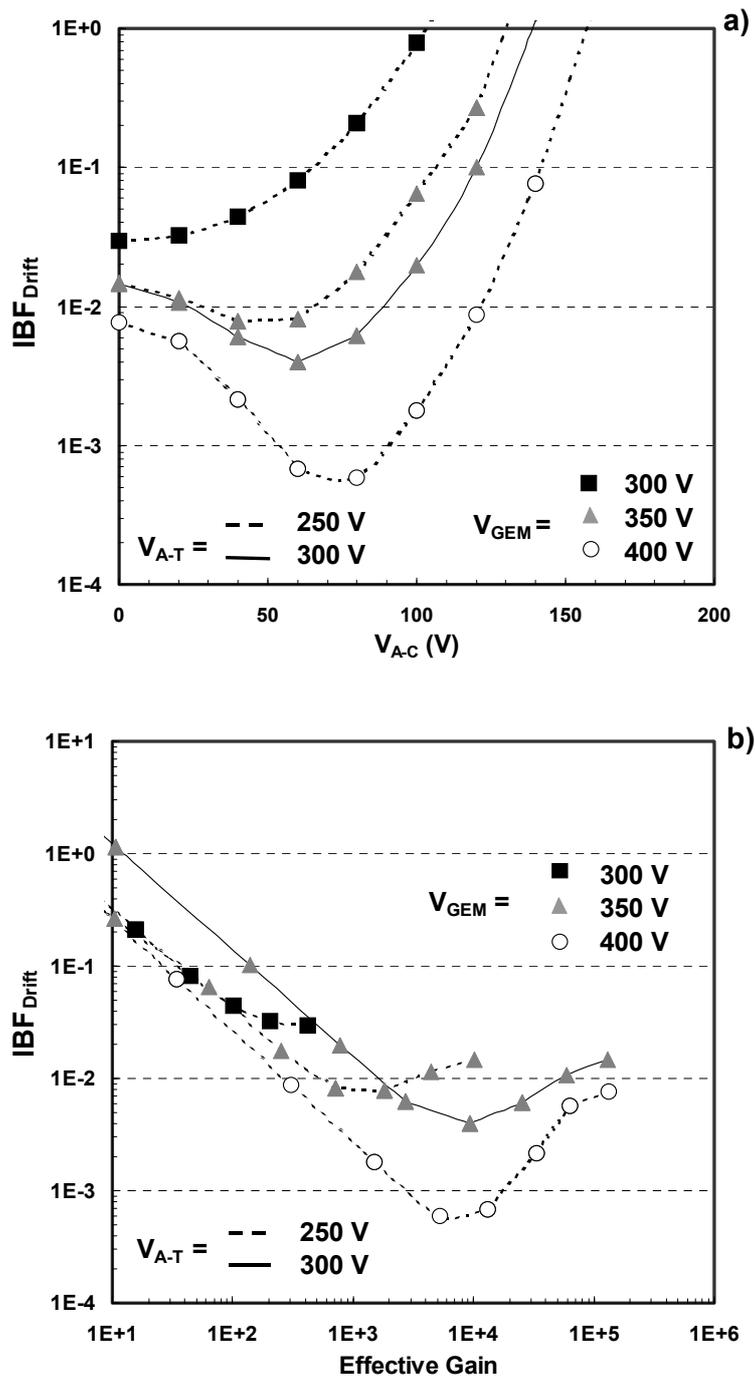

Fig.8 - Ion backflow to the drift region of the detector shown in fig. 7 as a function of $V_{A-C}$ (*a*), and of the total effective gain (*b*), for different $V_{A-T}$ and $V_{GEM}$ values; $E_{drift}$=0.5kV/cm; Ar/5%CH$_4$ at atmospheric pressure.